\documentclass[12pt]{article}
\usepackage[pdftex]{graphicx}
\usepackage{color}
\usepackage{amsmath}
\usepackage{amssymb}
\usepackage{euscript}
\usepackage{multirow}
\usepackage{color}
\usepackage{colortbl}
\usepackage{calc}
\usepackage{cite}

\newcommand{\figtxt}[1]{\footnotesize{#1}}

\makeatletter
\newlength\twolinebox@linelength
\newlength\twolinebox@columnheight
\newcommand{\twolinebox}[2]{%
   \setlength{\twolinebox@linelength}%
             {\maxof{\widthof{#1}}{\widthof{#2}}}%
   \setlength{\twolinebox@columnheight}{\heightof{#1}+\depthof{#1}+0.2em+0.4em/2+\heightof{0}/2}%
   \raisebox{0pt}[\twolinebox@columnheight][\heightof{\vbox{\vskip0.2em\hbox to 
   \twolinebox@linelength {#1\hfil}\vskip0.4em\hbox to 
   \twolinebox@linelength {#2\hfil}}}+\depthof{\vbox{\vskip0.2em\hbox to 
   \twolinebox@linelength {#1\hfil}\vskip0.4em\hbox to 
   \twolinebox@linelength {#2\hfil}}}-\twolinebox@columnheight+0.2em]{\vbox to 
   \twolinebox@columnheight{\vskip0.2em\hbox to 
   \twolinebox@linelength {#1\hfil}\vskip0.4em\hbox to 
   \twolinebox@linelength {#2\hfil}}}%
}

\newcommand{\tw}      {\textwidth}

\newcommand{\ie}[0]   {\textit{i.e.}}
\newcommand{\eg}[0]   {\textit{e.g.}}

\newcommand\MC@NLO[0] {\textsf{MC@NLO}}

\def\slashii#1{\setbox0=\hbox{$#1$}            
  \dimen0=\wd0                                 
  \setbox1=\hbox{\sl/} \dimen1=\wd1            
  \ifdim\dimen0>\dimen1                        
     \rlap{\hbox to \dimen0{\hfil\sl/\hfil}}   
     #1                                        
  \else                                        
     \rlap{\hbox to \dimen1{\hfil$#1$\hfil}}   
     \hbox{\sl/}                               
  \fi}

\definecolor{rltbrightred}{rgb}{1,0,0}
\definecolor{rltred}{rgb}{0.75,0,0}
\definecolor{rltdarkred}{rgb}{0.5,0,0}
\definecolor{rltbrightgreen}{rgb}{0,0.75,0}
\definecolor{rltgreen}{rgb}{0,0.5,0}
\definecolor{rltdarkgreen}{rgb}{0,0,0.25}
\definecolor{rltbrightblue}{rgb}{0,0,1}
\definecolor{rltblue}{rgb}{0,0,0.75}
\definecolor{rltdarkblue}{rgb}{0,0,0.5}
\definecolor{webred}{rgb}{0.5,.25,0}
\definecolor{webblue}{rgb}{0,0,0.75}
\definecolor{webgreen}{rgb}{0,0.5,0}
\definecolor{Black}{rgb}{0,0,0}
\definecolor{Greymax}{rgb}{0.65,0.65,0.65}
\definecolor{Greycen}{rgb}{0.75,0.75,0.75}
\definecolor{Greymin}{rgb}{0.85,0.85,0.85}
\definecolor{hl}{rgb}{
                0.909803922,       
                0.82745098,               
                0.909803922}

\begin{document}

\begin{titlepage}


\begin{center}
{\LARGE\bf On the contribution \\ \vspace{1mm}
of the double Drell--Yan process \\ \vspace{2mm}
to  $WW$ and $ZZ$  production  at the LHC$\,^{\star}$}
\end{center}

\vspace{7mm}

\begin{center}
{\large\bf  
 Mieczyslaw Witold Krasny$^{a}$ 
 {\rm and} 
 Wies\l{}aw P\l{}aczek$^{b}$}

\vspace{5mm}
{\em $^a$Laboratoire de Physique Nucl\'eaire et des Hautes \'Energies, \\
          Universit\'e Pierre et Marie Curie Paris 6, Universit\'e Paris Diderot Paris 7, \\
          CNRS--IN2P3, \\
          4 place Jussieu, 75252 Paris Cedex 05, France.}
\\  \vspace{2mm}
{\em $^b$Marian Smoluchowski Institute of Physics, Jagiellonian University,\\
         ul.\ \L{}ojasiewicza 11, 30-348 Krak\'ow, Poland.}
\end{center}

\vspace{18mm}
\begin{abstract}
\noindent  
In this paper, we investigate consequences of an assumption  that the discrepancy
of the predicted and observed $W^+W^-$ production cross sections at  the LHC
is caused by the missing contribution of the double Drell--Yan process (DDYP).
Using our simple model of DDYP~[{\it Acta~Phys.~Pol.~B} {\bf 45}, 71 (2014)],
we show that inclusion of this production mechanism  leads to   a satisfactory,
parameter-free  description of the two-lepton mass distribution for $0$-jet $W^+W^-$
events  and the four-lepton mass distribution for $ZZ$ events. In such a scenario,
the Higgs-boson contribution is no longer necessary to describe the data.
An experimental programme to prove or falsify such an assumption is proposed.
\end{abstract}

\vspace{7mm}
\begin{center}
{\it version published in    Acta Phys. Pol. B47 (2016) 4,1045 }
\end{center}

\vspace{20mm}
\footnoterule
\noindent
{\footnotesize
$^{\star}$The work is partly supported by the program of the French--Polish 
co-operation between IN2P3 and COPIN no.\ 05-116, 
and by the Polish National Centre of Science grant no.\ DEC-2011/03/B/ST2/00220.
}

\end{titlepage}

\section{Introduction}
\label{introduction}

The main motivation for our study presented in this paper was ``the $WW$ anomaly''
at the LHC, \ie deviations of the total cross sections for resonant $W^+W^-$ production
measured by the ATLAS~\cite{ATLAS:2012mec,ATLAS-CONF-2014-033} and CMS~\cite{Chatrchyan:2013yaa,Chatrchyan:2013oev}
experiments from their theoretical predictions, as shown Table~\ref{tab:WW-xsec}.  
The cross sections measured by ATLAS at the collision energies of 7~TeV
and 8~TeV are higher respectively by factors of $1.16$ and $1.22$ than the theoretical
predictions of the NLO QCD calculations obtained from
the {\sf MCFM} program~\cite{Campbell:2011bn}.
The corresponding factors for the CMS measurements are $1.11$ and $1.22$.

\begin{table}[htb]
\begin{center}
\begin{tabular}{||c||c|c||c|c||}
\hline\hline
  & {\bf ATLAS} & {\bf CMS} & \multicolumn{2} {c||}{{\bf Theory} $\sigma_{W^+W^-}$ [pb]} \\
\cline{4-5}
$\sqrt{s}$ & $\sigma_{W^+W^-}$ [pb] & $\sigma_{W^+W^-}$ [pb] &  ~~~ATLAS~~~ & CMS \\
\hline\hline
$7$ TeV & 
$51.9^{+2.0+3.9+2.0}_{-2.0-3.9-2.0}$ & 
$52.4^{+2.0+4.5+1.2}_{-2.0-4.5-1.2}$ & 
$44.7^{+2.0}_{-1.9}$ &
$47.0\pm 2.0$ \\
\hline
$8$ TeV & 
$71.4^{+1.2+5.0+2.2}_{-1.2-4.4-2.1}$ & 
$69.9^{+2.8+5.6+3.1}_{-2.8-5.6-3.1}$ & 
$58.7^{+3.0}_{-2.7}$ &
$57.3^{+2.3}_{-1.6}$ \\
\hline\hline
\end{tabular}
\caption[]
       {\figtxt{Values of the total cross section for resonant $W^+W^-$ production at the LHC
        measured by  the ATLAS and CMS experiments and predicted by the theoretical NLO QCD 
        calculations from the {\sf MCFM} program.}
        }
\end{center}
\label{tab:WW-xsec}
\end{table}%

Although each of these deviations does not exceed $2\sigma$, the fact that they are
present
in four measurements and all exhibit excess of data with respect to theory may indicate
that
there exist some extra processes contributing to the measured cross sections
that have not been taken
into account in the theoretical predictions. Such processes may have an important impact
on
the significance of the Higgs-boson signals at the LHC.

Recently, the NNLO calculations for the $W^+W^-$ production have been published~\cite{Gehrmann:2014fva}.
They predict increase of the theoretical cross sections at these energies by about $9\%$
with respect
to the NLO results, so they get closer to the experimental measurements but do not remove
the
differences completely. These calculations have been done, however, for the total cross
sections  only
and not for distributions considered by the LHC collaborations in the Higgs searches.
The question if the increase of the cross section is in the Higgs-signal region or in the
Higgs-monitoring region, or both, which is critical to our analysis, remains thus open.
Therefore, we shall not use them in this paper.
We can do this in the future when the ATLAS and CMS collaborations apply them in their
data analyses.

Since the observed $W^+W^-$ cross section discrepancies  at  8~TeV are by a factor of
$\sim 3$ higher than the expected
contribution coming from the Standard Model (SM) Higgs-bosons decays,  special
measurement procedures were used by the ATLAS and CMS collaborations,
see \eg\ Refs.~\cite{ATLAS:2014aga,Chatrchyan:2013iaa}, to increase
their  sensitivities to the Higgs signal.
The normalisation of the predicted background was rescaled to fit the data in the
monitoring region,
where the Higgs contribution is negligible.
Then, the rescaled background was used in the Higgs-search region to determine the strength
of
the Higgs signal.

In the case of the ATLAS experiment, the normalisation of the theoretical predictions
for the background was multiplied by a factor of $1.22$ for the 8~TeV analysis  of the
$0$-jet $e \mu$ events~\cite{ATLAS:2014aga},  where the sensitivity to the Higgs boson
decays is the highest.
This value is compatible with the discrepancy  of the measured and predicted total cross
sections
for the resonant $W^+W^-$ production at 8~TeV, see Table~\ref{tab:WW-xsec}.
The actual value of the rescaling factor for the CMS analysis is not explicitly quoted in
the
their papers. Therefore, the CMS data will not be used
in the presented analysis\footnote{A more general comment is obligatory for discerning
unavoidable caveats of the analysis  presented in this paper.
The LHC collaborations publish very rarely their detector-effect unfolded distributions.
The Higgs papers are no exceptions here.
This preempts a fully  irrefutable justification of any external analysis of these
distributions,
including the one   presented in this paper. All we can do is
to  try  to minimise the impact of the necessary underlying assumptions on the event
selection
efficiencies and detector smearing effects. Their remaining impact can be evaluated
only by the relevant collaborations.}.

In this paper, we shall make a bold {\it assumption}
that the missing process accounting for the cross-section discrepancy,
not considered so far in the calculations of the theoretical cross sections,
is the  double Drell--Yan process (DDYP) resulting from double-parton scattering (DPS)
in proton--proton collisions.
It should be stressed that the strength of DDYP can, at present, be neither directly
constrained by experimental data nor predicted theoretically.

If one includes DDYP as a contributor to the $W^+W^-$ production processes,
it is bound to contribute as well  to the $ZZ$ production processes with a fully
constrained strength.
The immediate  question which may be asked is if after adding the DDYP contribution
to the Higgs-boson searches background, both in the
$W^+W^-$ and $ZZ$ channels, there would still be a need to include the contribution coming
from the Higgs-boson decays or, putting it alternatively,   to which extent the DDYP
contribution,
on top of curing the $WW$ anomaly,
could mimic the Higgs-boson signal  in the 125~GeV Higgs-sensitive phase-space regions.

To answer this question, we present a  coherent analysis of the DDYP contribution to the
$H \rightarrow W^+W^-$ and $H \rightarrow ZZ^*$ decay-channel backgrounds.
We use the data corresponding to the total integrated luminosity collected so far at the
LHC.
Once the overall normalisation of the DDYP contribution is fixed
to explain the $WW$ anomaly,  our DDYP model
does not have any more free parameters, thus can be easily falsified
by comparing its predictions to experimental data.

The DDYP contribution to the Higgs searches background in the  $H \rightarrow ZZ^*$ decay
channel was already analysed in our previous work~\cite{Krasny:2013aca}.
Using a  simplified model of DDYP,  we have demonstrated the appearance
of a  peak in the four-lepton invariant  mass, $m_{4l}$,  distribution in the $\sim 125$~GeV
Higgs-signal region.
This  ``Higgs-like'' peak is not driven by the details of the DDYP model. It is generated
by the  interplay of a steeply falling $m_{4l}$
distribution  and the kinematical  threshold effect driven by  the experimental cuts on
the outgoing leptons variables:
the minimal transverse momenta of leptons and the minimal invariant mass of the opposite
charge lepton pairs.
These cuts are similar in the ATLAS and CMS analyses, and therefore result in similar DDYP
peak positions within a 2~GeV interval.

It has to be stressed that the claim of  the ATLAS and CMS collaborations that DDYP can be
neglected as a potentially alarming source of background was based on the
assumption of uncorrelated: (1) longitudinal momentum, (2) transverse position, (3)
flavour, (4) charge and (5) spin
of the partons taking part in DDYP,  and on the  assumption  of the process-independent
value of
$\sigma_{\rm eff}$,  governing the strength of  the DPS processes.
The above, in our view unjustified,  assumptions  lead  to a significant underestimation
of the contribution
of DDYP to the Higgs searches background. As we argued in Ref.~\cite{Krasny:2013aca}, its
contribution must be, given the lack of the
adequate theoretical calculations,  determined experimentally,
\eg\ by using the experimental methods proposed therein and in Ref.~\cite{Krasny:2011dj}.

There are three main reasons for writing this paper:
\begin{enumerate}
\item
In Ref.~\cite{Krasny:2013aca}, we applied our DDYP model only to the ATLAS data for the
$ZZ$ channel, while in this paper, we apply it to the  ATLAS data for the $WW$ channel, and
after fitting the normalisation factor, we use~it, parameter-free, in the $ZZ$ channel.
\item
The analysis in Ref.~\cite{Krasny:2013aca} was done only for the partial ATLAS data
available at that time, here we use the full data collected by ATLAS in the LHC Run~1 to
check if
our DDYP model can still describe these data.
\item
We propose new experimental methods of testing DDYP at the LHC, see
Section~\ref{sec:discussion}.
\end{enumerate}

The paper is organised as follows. In the next section, we present numerical results of our
analysis. Section~\ref{sec:discussion} includes a discussion of the results as well as a general
discussion of the interplay between the DDYP and gluon--gluon scattering
contributions. It contains a proposal of measurements capable to
elucidate the role of DDYP in $WW$ and $ZZ$ production processes.
Finally, Section~\ref{sec4} concludes our paper.

\section{Results}
\label{sec:results}

The numerical results presented below have been obtained using the Monte Carlo event
generator {\sf WINHAC}~\cite{WINHAC:MC,Placzek2003zg,Placzek:2013moa} with the same model of DDYP
and the same input parameters as in our previous paper~\cite{Krasny:2013aca}.

The starting point to our quantitative studies in this paper is a shape analysis of the
two-lepton
mass, $m_{ll}$, distribution and its fit to the ``$e\mu\nu\nu + 0$-jets'' final-state data
presented
by the ATLAS Collaboration in a wide $m_{ll}$ range for the full data statistics at
8~TeV in Ref.~\cite{Aad:2013wqa}.

\begin{figure}[!ht] 
  \begin{center}
    \includegraphics[width=1.05\tw]{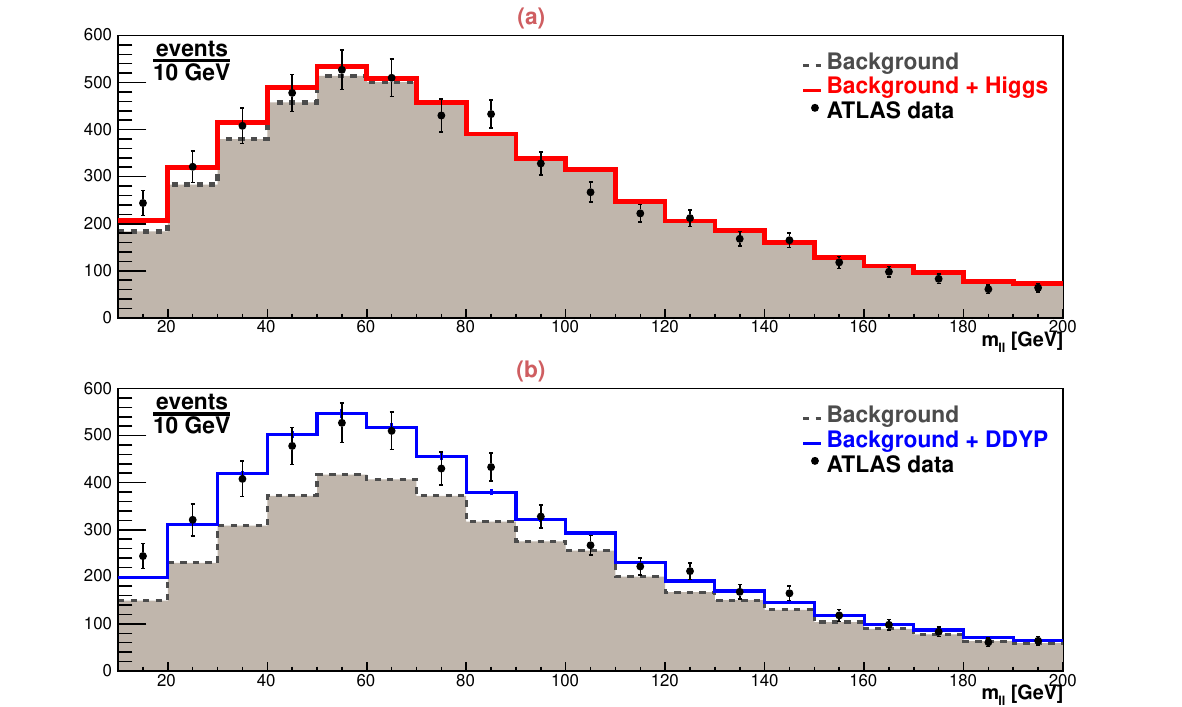}
    \caption[]
       {\figtxt{The $m_{ll}$ distributions for the $H\rightarrow W^+W^-$ searches at
                  $\sqrt{s} = 8$~TeV: (a) the rescaled (ATLAS) background without (grey histogram)
                  and with (solid/red line) the SM Higgs boson contribution compared with the ATLAS
                  data (black dots)~\cite{Aad:2013wqa}, and (b) the canonical background without
                  (grey histogram) and with (solid/blue line) the DDYP contribution compared with
                  the same ATLAS data (black dots).}
        }
       \label{fig:WW}
  \end{center}
\end{figure}

In Fig.~\ref{fig:WW}~(a), we present the results of the ATLAS Collaboration of
Ref.~\cite{Aad:2013wqa}
for the collision energy of 8~TeV. The background and the Higgs decays contributions
are shown separately. This plot reflects the necessity of including the Higgs contribution
to obtain a satisfactory description of the $m_{ll}$ distribution  in the region of its
small values,
where the sensitivity to the 125~GeV Higgs boson is the highest.
However, we would like to stress again that the original background prediction was, in this
case, rescaled
by the ATLAS Collaboration by the factor of $1.22$ in order to get a good description of
data in
the monitoring region.

For the plot presented in  Fig.~\ref{fig:WW}~(b), we re-normalise back the background
distribution to its
original, canonical  theoretical predictions
by dividing the background shown in Fig.~\ref{fig:WW}~(a) by a factor of $1.22$, and
subsequently add the DDYP contribution with a  normalisation of its prediction determined
by
minimisation of the   $\chi^2/{\rm d.o.f.}$ in the fit of the sum of the DDYP and theoretical
background
contributions to the data.
The Higgs contribution is no longer necessary to obtain a satisfactory description of the
data by the canonical background model if the DDYP contribution is included.

Indeed, the values of $\chi^2/{\rm d.o.f.}$ corresponding to the $m_{ll}$ distributions
in Fig.~\ref{fig:WW}
are $1.2$ for the ``rescaled background plus the SM Higgs'' scenario (with the $p$-value
equal to $0.26$)
and $0.8$ for the ``canonical background plus the DDYP'' scenario (with $p$-value equal to
$0.70$).
Both descriptions of the data are acceptable on a statistical basis, although the obtained
values
of the likelihood test give some better preference to the ``DDYP plus  canonical
background'' predictions.
We stress again that in the latter case, the theoretical predictions of the SM background
do not need to be rescaled to describe the data, as it was  done in
Refs.~\cite{ATLAS:2014aga,Chatrchyan:2013iaa}.

\begin{figure}[!ht] 
  \begin{center}
    \includegraphics[width=1.05\tw]{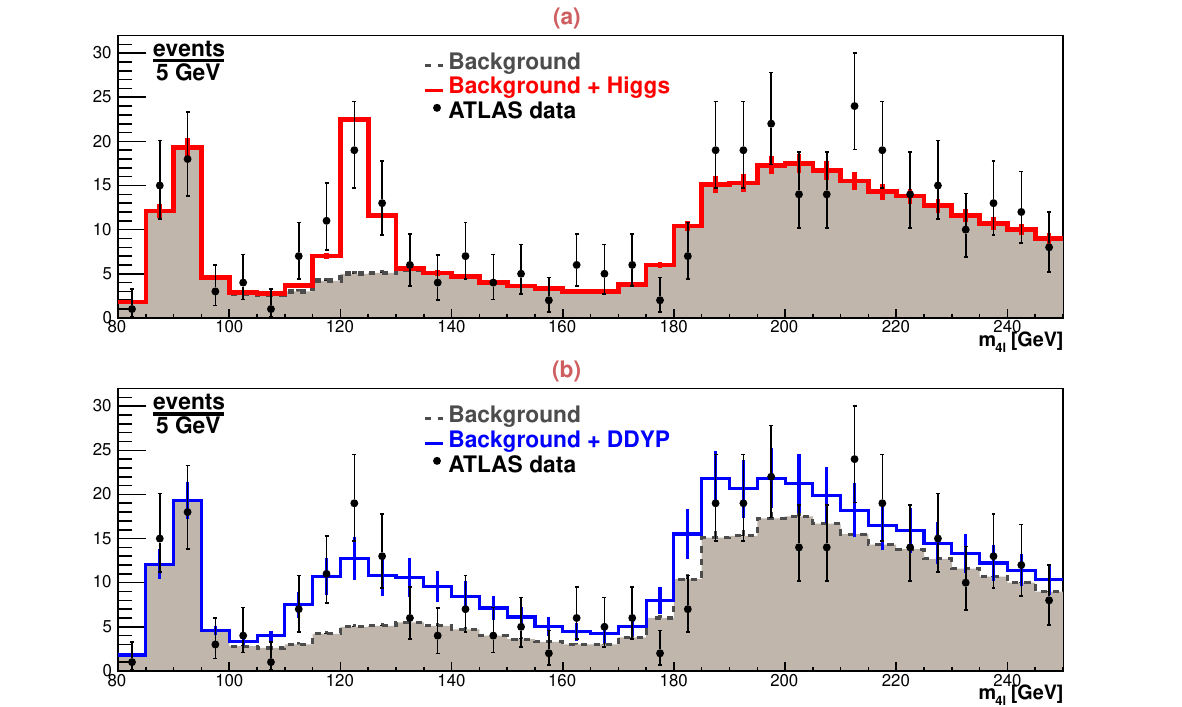}
    \caption[]
       {\figtxt{The $m_{4l}$ distributions for the $H\rightarrow ZZ$ searches at
                  $\sqrt{s} = 8\,$TeV: (a) without (grey histogram) and with (solid/red line)
                   the SM Higgs-boson contribution compared with the ATLAS data (black dots)~\cite{Aad:2013wqa},
                  and (b) the (ATLAS) background plus the DDYP contribution
                  (solid/blue line) compared with the same ATLAS data (black dots).}
        }
       \label{fig:ZZ}
  \end{center}
\end{figure}

Having determined the normalisation of DDYP cross section using the $W^+W^-$ channel,
we can now study the corresponding DDYP contribution to the $m_{4l}$ distribution
for the $ZZ$ channel. In our model, the dominant contribution to the DDYP cross section
comes form the $q\bar{q}$ excitations of the proton sea. Thus,  the relative normalisation
of the DDYP contributions for the  $W^+W^-$  and $ZZ$ channels is driven, on the modelling
side,
only by the relative strength of the $u\bar{u}$ and $d\bar{d}$ excitations, reflecting
the ratio  of the $\bar{u}$ and $\bar{d}$ PDFs (for more details, see
Ref.~\cite{Krasny:2013aca}).
On the experimental side,  it reflects the relative efficiencies of selection of $W^+W^-$
and $ZZ$ events in their kinematical acceptance regions\footnote{For the analysis of the
$W^+W^-$ and $ZZ$ spectra, we select
events according to the selection criteria  used for the Higgs searches in the
corresponding
decay channels. We assume the same efficiency of the $W^+W^-$ and $ZZ$ event
selection and neglect the detector smearing effects.}.
The DDYP model predictions for the $ZZ$ channel are thus fully constrained.

Again, as for the $W^+W^-$ channel,  the ATLAS data at 8~TeV~\cite{Aad:2013wqa} are
presented first in Fig.~\ref{fig:ZZ}~(a)  with the SM Higgs contribution and the canonical
background contributions shown separately.
In Fig.~\ref{fig:ZZ}~(b), we replace the Higgs contribution by the parameter-free DDYP model
predictions.
The values of $\chi^2/{\rm d.o.f.}$ corresponding to $m_{4l}$ distributions in
Fig.~\ref{fig:ZZ}
are $0.65$ for the background plus SM Higgs scenario (with the $p$-value equal to $0.94$)
and $0.95$ for the background plus our DDYP predictions (with $p$-value equal to $0.55$).
Both descriptions of data are statistically acceptable, however in this case, the SM Higgs
prediction
is slightly more preferred.

In contrast to the SM Higgs scenario, DDYP contributes not only to
the $m_{4l}$ range near 125~GeV but also above the threshold of the on-shell $ZZ$
production.
We have calculated that with $W^+W^-$-fixed  normalisation, DDYP will result in $31$ extra
events
in the range of 190~GeV~$\leq m_{4l} \leq 250$~GeV, which agrees at the
$1\sigma$ level
with the ATLAS data excess of $20\pm 13.5$ events over  the SM background~\cite{Aad:2013wqa}.
CMS also observes an excess of the data with respect to the background in this kinematical
region
at a comparable level of $\sim 13\%$~\cite{Chatrchyan:2013mxa}.

\section{Discussion}
\label{sec:discussion}

Our numerical results presented in the previous section show that once the DDYP
contribution
is normalised to account  for the missing contribution to the $W^+W^-$ cross section,
one obtains a satisfactory  predictions for  the excess of events
observed by the ATLAS experiment both in the $m_{ll}$ and $m_{4l}$ distributions.
These  excesses of events were  attributed by the ATLAS and CMS
collaborations  to the  SM Higgs-boson decays to $ZZ$ and $W^+W^-$
with $m_H \approx 125$~GeV.

The presented results correspond to  the two extreme scenarios
explaining the excess of the experimental data over the SM background:
the ``SM-Higgs-only'' and  the ``DDYP-only''.
There is, of course, a possibility that the two could contribute together.
For example, in the case of the $m_{4l}$ distribution, the SM Higgs contribution
could better describe the peak width in the data in the 125~GeV region, while DDYP can
add some
events in this region but also in the higher mass range. This would lead to a different
values
of the Higgs couplings to the $W$ and $Z$ bosons with respect to those quoted in the
canonical ATLAS and CMS analyses which neglect the DDYP contribution altogether.

The intriguing question is if the DDYP contribution  could be sufficiently large to mimic
fully the Higgs
signal at the LHC. An equally important  question is why it was not considered
as important  background source in the data analyses by the ATLAS and CMS collaborations.
Let us start with the answer to the latter question before addressing the former one.

The DDYP contributions  were studied by both collaborations with the help of the standard
parton shower Monte Carlo generators and were found to be negligible.
The reason why these generators predict such small cross sections for DDYP is related to
the assumption of a completely uncorrelated two single-parton scattering (SPS)
modelling of two Drell--Yan processes in a single proton--proton collision.
The canonical DPS models assume the value of the so-called effective double-parton
scattering
cross section: $\sigma_{\rm eff} \approx  15$~mb.
This parameter normalises the DPS cross section with respect to the product of two SPS
cross sections,
see \eg\ \cite{Krasny:2013aca}.

The above value has been measured, but only for the cases of DPS involving at least one
gluon in a pair of colliding partons.
There are at least three reasons why the value of $\sigma_{\rm eff}$
may be significantly lower for
the same flavour, opposite-charge and spin quark--antiquark pairs relevant to the
$ZZ$ and $W^+W^-$ production processes:
\begin{itemize}
\item
The quark--antiquark excitations of the proton involves partons of the same flavour.
For example, if an $s$-quark takes part in the production of one of the vector boson,
its $\bar{s}$ partner is already present in the wave function of the colliding proton.
This enhances by a large factor, with respect to the canonical picture in which
the PDFs of both partons are considered as uncorrelated,
the probability that it may take part in the production of the second vector boson.

\item
The transverse--plane correlation length for the same flavour and opposite charge
$q\bar{q}$ pairs
is significantly smaller than for the gluon--gluon pairs because of the dominance of the
local charge
and flavour conserving proton excitations --- note that at the hardness scale of the
vector-meson pair production processes and for the $x$-region relevant to the
Higgs searches, there is less than one fixed-flavour  $q\bar{q}$-pair excitation per
incoming proton,
while there is more than $100$ gluon pairs covering uniformly, glueing together,  the
proton volume.
No doubt, the strength of  the gluonic DPS must reflect the transverse size of the proton,
while DDYP should reflect rather the typical size of a quark--antiquark dipole.

\item
The long-lived $q\bar{q}$ excitations within  the proton which do not lead to the
significant
effects modifying its overall spin involve quarks
and antiquarks of opposite spins. Collisions  of such pairs produce always
vector-boson pairs with the total spin equal to zero, as in the case of the Higgs-boson
decays.
\end{itemize}

All the above amplification effects cannot  be calculated within the pres-ently available
QCD perturbative
methods. This, however,  does not mean that they do not exist.
The LHC provides the unique opportunity to measure
the respective quark--antiquark flavour, transverse plane, longitudinal momentum and spin
 correlations in the proton.
The $XY$-pair production processes, where $X,Y \in \{\gamma^*,Z,W, J/{\mit\Psi},{\mit\Upsilon}\}$,
provide an excellent experimental
testing ground for the quark--antiquark correlation models, in particular~if:
(1) data are collected at two or more collider energy settings (allowing
to separate experimentally the quark and gluon originating processes~\cite{Krasny:2011dj}),
(2) the data are collected not only in the $pp$ but also in the $pA$ and $AA$
collision modes (to control the transverse--plane parton correlations) and, most
importantly,
(3) the DDYP effects  are measured both in the Higgs-signal  and  monitoring regions.
In addition, the measurement of the relative
strengths of these processes provides a clear experimental test of the robustness
of the SM Higgs interpretation of the data with respect to alternative mechanisms of
the electroweak symmetry breaking.
As long as such experimental studies are not made, %
the DDYP model should be, in our view, considered as equally plausible as the Higgs model
in explaining the source of the excesses of events in the $ZZ$ and $W^+W^-$ channels
observed at the LHC.

In this paper, we avoid giving any value of  $\sigma_{\rm eff}$ for our DDYP,
since, in general, DPS does not simply factorise into the product of two SPS processes,
especially when the DPS contribution is sizeable, as it is in our case;
examples of that are shown \eg\ in Ref.~\cite{Golec-Biernat:2014nsa}.
Even if it were possible, the ATLAS and CMS experiments do not provide the necessary
information
(detector efficiencies and smearing effects) to translate our normalisation factor to
$\sigma_{\rm eff}$.
At the generator level, this normalisation factor is of the same order of magnitude as
in Ref.~\cite{Krasny:2013aca}.

Finally, let us address a more subtle question of the interplay between DDYP and
the electroweak-boson pair production in the gluon--gluon fusion process.
The latter  processes are included in the {\sf MCFM} program~\cite{Campbell:2011bn} as
well as in
the {\sf gg2WW}~\cite{Binoth:2006mf} and {\sf gg2ZZ}~\cite{Binoth:2008pr} generators
which are used by the ATLAS and CMS collaborations for theoretical predictions of the
respective
SM background to the Higgs signal in its $WW$ and $ZZ$ decay channels.
These calculations include also the so-called crossed-box contribution in which two
incoming gluons
split into quark--antiquark pairs and then a quark (antiquark) of one pair interacts with
an antiquark (quark) of the second pair, leading to DDYP. One may, therefore, think that
the processes we consider in this paper are already  included in the theoretical
predictions
of the SM background to the respective Higgs signals.

In our opinion, the gluon--gluon fusion calculations include only partially the DDYP
contribution
and may even underestimate the included crossed-box diagram effects.
Firstly, they take into account
only on-shell incoming gluons which are purely left- or right-handed, while, as we argue in
Ref.~\cite{Krasny:2013aca},  one should consider all possible gluon polarisations.
For the spin-zero vector boson pairs mimicking the Higgs signal, a particular care must
be given to
the full set of processes producing on-shell and off-shell $q\bar{q}$ pairs with
compensating spins.
Such pairs may be produced by longitudinally polarised gluons but  also by  the
higher-twist effects.
To our best knowledge these effects are not included in the existing theoretical
calculations.

As was shown in a detail in Ref.~\cite{Gaunt:2011xd},
the crossed-box contributions exhibit a collinear singularity when the spins of the
incoming on-shell gluons
sum to zero. The typical collinear singularity is,  within the leading-twist approach,
damped in this case to the logarithmic
(integrable) singularity due to the vector structure of QCD for the on-shell initial-state
gluons,
\ie when they are purely left- or right-handed~\cite{Gaunt:2011xd}.
Off-shellness of longitudinally polarised gluons should,
to some extent, damp the singularity, but will this reduce the enhancement due to the
collinear quark--antiquark pair emission?

The question which remains to be answered is not only how frequently
the incoming gluons are longitudinally polarised
but also  what is the probability for the $q\bar{q}$ pair propagating over large
distances,
before annihilating into a vector-boson pair,
to become  a spin-zero pair due to soft colour interactions with the medium.
In both cases, the total spin of the colliding  $q\bar{q}$ dipoles is zero,  leading to
a spin zero configuration of the  final  vector-boson pairs.

Another   problematic issue is the question of the renormalisation and factorisation
scales.
In typical fixed order
QCD calculations, as given in
Refs.~\cite{Campbell:2011bn,Binoth:2006mf,Binoth:2008pr,Gaunt:2011xd}, these scales are
set equal to each other and taken as a hard-process scale.
In the case of the processes under consideration, this scale is of the order of $\sim
100$~GeV.
Is using such a high scale justified for processes in which incoming gluons split into
almost
collinear quark--antiquark pairs? In our opinion it is not. In such cases, a better choice
for
the argument of the running QCD coupling may  be not the hard process scale
but rather the transverse momentum of the emitted quarks, see \eg\
Ref.~\cite{Amati:1980ch}.
Generally, this kind of a scale is used in popular QCD parton shower generators.
This is, however, often not the case in the fixed-order QCD calculations.
The value of the running $\alpha_\mathrm{s}$ between the scales of $\sim 100$~GeV and $\sim
1$~GeV
 is increased by a factor of $\sim 5$.
Since the considered process involves $\alpha_\mathrm{s}^2$, we can easily get the enhancement
factor of
$\sim 25$. The factorisation scale for such collinear splittings should also be set to a
similar value.

The effects discussed above should lead to enhancements in low $p_\mathrm{T}$ regions of produced
electroweak bosons. This is hard to observe  in the case of the $W^+W^-$ production, as
the
transverse momenta of neutrinos from the leptonic $W$-boson decays cannot be measured
individually,  but could be seen rather easily in
the $ZZ$/$Z\gamma^*$ processes through their charged lepton decay channels.
Therefore, the latter processes can provide an important test of the
interplay between the DDYP and gluon--gluon scattering  contributions.

\section{Conclusions}
\label{sec4}

In this paper, we have investigated the consequences of the hypothesis that the
double Drell--Yan process (DDYP) accounts for the excess of the measured $W^+W^-$
cross section with respect to its theoretically predicted value.
This assumption determines the absolute normalisation of the predictions of our
simple DDYP model introduced in Ref.~\cite{Krasny:2013aca}.
This normalisation factor is the only free parameter of the model.

We have demonstrated that adding the above absolutely normalised DDYP contribution
to the canonical SM background is sufficient for
a satisfactory description of the $m_{4l}$ spectra for the $ZZ$ final state
and $m_{2l}$ spectra for the $W^+W^-$  final state, with a comparable
fit quality as in the model assuming the existence of the SM Higgs boson.

We have argued that the DDYP contributions may indeed be significantly larger
than that expected in the naive canonical DPS models because of the
strong charge, flavour, longitudinal momentum, transverse position and spin
correlations of the quark--antiquark pairs  participating in DDYP.
Such a possibility has not been excluded so far, neither by the theoretical
calculations nor by the experimental measurements.
We have presented some theoretical arguments in favour of such a DDYP contribution
and proposed  a measurement programme to  test it experimentally at the LHC.

\vspace{5mm}
\noindent
{\large\bf Acknowledgements}
\vspace{3mm}

\noindent
We would like to thank S.P.~Baranov and J.R.~Gaunt for useful discussions.
We are indebted to the Academic Computer Centre CYFRONET AGH in Krak\'ow, Poland, where our
numerical simulations were performed with the use of the computing cluster {\sf Zeus}.

\vspace{5mm}
\noindent


\end{document}